%% file: main.tex
\documentclass[fleqn]{wlscirep}
\usepackage[utf8]{inputenc}
\usepackage{graphicx}
\usepackage{color}
\usepackage[normalem]{ulem}
\usepackage{simplewick}
\usepackage{xspace}
\usepackage{pgf,tikz}
\usetikzlibrary{arrows,intersections,decorations,decorations.markings,backgrounds,calc,patterns}
\usepackage{rotating}

\newcommand{\ie}{i.\,e.\@\xspace}

\include{journals}

\title{Relevance of topological disorder on the directed percolation phase transition}

\author[1*]{Manuel Schrauth}
\author[1]{Jefferson S. E. Portela}
\author[1]{Florian Goth}
\affil[1]{Institute of Theoretical Physics and Astrophysics,	University of W\"urzburg, 97074 W\"urzburg, Germany}
\affil[*]{manuel.schrauth@uni-wuerzburg.de}

\begin{abstract}
	Despite decades of research, the precise role of topological disorder in critical phenomena has yet to be fully understood. A major contribution has been the work by Barghathi and Vojta~\cite{barghathi2014}, which uses spatial correlations to explain puzzling earlier results. 
	However, due to its reliance on coordination number fluctuations, their criterion cannot be applied to constant-coordination lattices, raising the question, for which classes of transitions this type of disorder can be a relevant perturbation. In order to cast light on this question, we investigate the non-equilibrium phase transition of the two-dimensional contact process on different types of spatial random graphs with varying or constant coordination number. Using large-scale numerical simulations, we find the disorder to be relevant, as the dynamical scaling behaviour turns out to be non-conventional, thus ruling out the directed percolation universality class. We conjecture that the relevance of topological disorder is linked to how strongly connected the lattice is.
	Based on this assumption, we design two analysis tools that succeed in qualitatively distinguishing relevant from non-relevant topological disorders, supporting our conjecture and possibly pointing the way to a more complete relevance criterion.
\end{abstract}

\begin{document}
	
\definecolor{zzttqq}{rgb}{0.6,0.2,0}
\definecolor{ffqqtt}{rgb}{1,0,0.2}

\flushbottom
\maketitle
\thispagestyle{empty}

\input{intro}
\input{definitions}

\input{results}
\input{discussion}

\input{conclusion}

\section*{Acknowledgments}
We thank H.~Hinrichsen for helpful discussions. M.S.~thanks the Studienstiftung des deutschen Volkes for financial support. This work is part of the DFG research project Hi~744/9-1. F.G. thanks the DFG for funding through the SFB  1170 “Tocotronics” under the grant number Z03.
We are grateful to the Rechenzentrum Würzburg for providing computational resources through the JULIA Cluster.

\section*{Additional information}
\textbf{Competing interests:} The authors declare no competing interests.\\
\textbf{Author contributions statement:} M.S. and J.S.E.P. conceived the project, and all authors contributed to producing the code, conducting the simulations, analysing the results, and writing the manuscript.

\input{appendix}



{\small \input{references.bbl} }

\end{document}

%% file: journals.tex
\def\physicaA{Physica A}

%% file: intro.tex
\section*{Introduction}
\label{sec:Introduction}

	Continuous phase transitions in many-particle systems are characterized by long-range collective behaviour of microscopic degrees of freedom, which arises close to a specific value of the control parameter, the so-called critical point. A prominent example is the contact process (CP), where the ratio between offspring creation and spontaneous annihilation controls whether the system can maintain a finite number of active particles or eventually reaches an empty absorbing state~\cite{hinrichsen2000}. Systems exhibiting a phase transition can be categorized into universality classes, showing the same macroscopic behaviour despite differences on certain microscopic characteristics of the system, such as the precise geometry of the lattice. The contact process is one particular realization of the directed percolation (DP) universality class, which collects all non-equilibrium absorbing-state phase transitions with a scalar order parameter, local interactions and no additional symmetries or conservation laws~\cite{janssen1981,grassberger1982}.
	
	Quenched spatial disorder, however, introduces an external perturbation which may, in general, change the character of the phase transition. In the case of uncorrelated disorder, such as impurities in a regular lattice, Harris~\cite{harris1974} argued in the derivation of his famous criterion that locally varying coordination numbers lead to fluctuating local transition points. Averaging these fluctuations over the correlation volume and comparing them with the absolute distance from criticality, disorder is found to become asymptotically irrelevant whenever $d\nu>2$, where $ \nu $ denotes the correlation length exponent and $d$ is the dimension of the system. For the DP universality class, Harris' inequality is violated for two and three dimensions ($ \nu\approx0.73,\,\,0.58 $), entailing that uncorrelated disorder is relevant, \ie alters the transition. In fact, a series of numerical simulations of the contact process on diluted lattices~\cite{vojta2005,vojta2009,wada2017}, as well as analytical RG methods~\cite{hoyos2008,igloi2005,igloi2018}, showed that these disordered systems belong to the random transverse-field Ising model (RTFIM) universality class. As a consequence, the dynamical (time-dependent) observables display an infinite-randomness critical point with slow activated scaling, rather than conventional power-law scaling.
	
	Despite its explanatory power, the Harris criterion is not able to account for some of the results for \emph{topologically} disordered structures. Most strikingly, the 3D Ising model and the 2D contact process on Voronoi-Delaunay (VD) triangulations generated from random point clouds have been found to show clean universal behaviour~\cite{janke2002,lima2008,oliveira2008,oliveira2016}, even though the Harris inequality is violated in both cases~\cite{odor2008,henkel2009,pelissetto2002}. As a consequence, Barghathi and Vojta put forward an improved criterion~\cite{barghathi2014} (HBV). By taking spatial correlations into account, it is able to explain the unchanged universality in terms of a fast decay of disorder fluctuations with diverging correlation length as the critical point is approached. Specifically, they found  that quenched topological disorder is irrelevant in systems that satisfy $ a\nu >1 $, where $ a $ is the dimension-dependent \emph{disorder decay exponent} defined in terms of a spatial block coarse-graining procedure of lattice coordination numbers. It should be emphasized that $a$ is a geometrical property of the lattice rather than a property of the physical model. In Ref.~\citen{barghathi2014} it is shown that in 2D, $a\!=\!1$ for randomly diluted regular lattices, whereas $a=3/2$ for VD lattices, rendering disorder of the VD type less relevant than uncorrelated disorder and hence clarifying some of the results which could not be explained by the original Harris criterion. However, the HBV criterion also fails to provide a complete description, as we have recently shown~\cite{schrauth2018b} using a VD$^+$ lattice, which consists in a VD triangulation with additional random local bonds. Despite the additional uncorrelated, slowly decaying, coordination number disorder ($a=1$), the CP phase transition on this lattice unambiguously belongs to the clean DP universality class. 
	
	There are also lattices inaccessible to the HBV criterion. The criterion takes fluctuations in the coordination number as the source of local transition fluctuations, which in turn determine the relevance of disorder perturbations; however, for lattices where such fluctuations are absent, $a$ is not defined. A simple example is the Voronoi graph (VG) where every site has exactly three neighbours, as can easily be seen from its dual being the VD \emph{triangulation}. Another example two of us recently introduced~\cite{schrauth2018a,paper-cc} is the Constant Coordination (CC) lattice, obtained from random bonds made local through dynamical rewiring.
	Numerical simulations of the 2D Ising model on those two lattices with constant coordination number (VG and CC) produce rather distinct results. Whereas for the VG, the system remains in the clean Ising universality class~\cite{paper-vg}, the CC lattice shows varying exponents and hence no clear evidence of clean universal behaviour~\cite{schrauth2018a}. In fact, the latter results are qualitatively similar to those on diluted regular lattices which, owing to the ambiguity of their universality character, received considerable attention through the last three decades. Specifically, deviations of the system's critical exponents from their clean values have been explained both as a disorder-dependent non-universal behaviour~\cite{kim1994a,kim1994b,selke1994b,ziegler1994,kuehn1994,kim2000} (the so-called weakly universal scenario), and as resulting from strong logarithmic corrections due to the marginality of the model with respect to the Harris criterion (strong universality scenario)~\cite{dotsenko1981,jug1983,shankar1987,shalaev1994,ludwig1987,martins2007,fytas2010,fytas2013,zhu2015}.
	
	This marginality makes it difficult to tell whether disorder of the CC type ultimately represents a relevant perturbation. In this paper, we therefore turn our attention to a 2D contact process, as it should allow for a clearer distinction between universal and non-universal behaviours, due to its smaller correlation length exponent, $\nu\approx0.73$. We find that the CC lattice disorder is a relevant perturbation, as the dynamics shifts from the conventional power-law scaling into the activated scaling regime. It turns out that the results are much like those for the contact process on Gabriel and Relative Neighbour Graphs, also investigated in this work, for which the HBV criterion already predicts a failure of clean universal behaviour. This raises the question of why disorder of the CC type is more relevant than that of the Voronoi type, even though both share the absence of the coordination number fluctuations. We make a case for relevant disorders to be those which lead to low connectivity, and present our contribution to answering this question in the remainder of this work, which is organised as follows: First we introduce the considered lattices and the physical model (Sec.~\ref{sec:Lattices}), followed by the CP results and discussion (Sec.~\ref{sec:Results} and \ref{sec:Discussion}) and, finally, our concluding remarks (Sec.~\ref{sec:Conclusion}).

%% file: definitions.tex
\section*{Definitions}
\label{sec:Lattices}

\subsection*{Constant Coordination Lattice}
\label{sec:ConstantCoordinationLattice}
	
	Instrumental for studying the relevance of coordination number fluctuations are lattices where such fluctuations are absent. One such lattice is the Constant Coordination (CC) lattice~\cite{schrauth2018a,paper-cc}, where by construction every site is locally connected to exactly $q$ other sites. A sample with $q=4$ (CC4) can be seen in Fig.~\ref{fig:lattice_overview}. Starting from a cloud of randomly distributed sites, bonds between random pairs of sites are gradually introduced until each site has exactly $ q $ neighbours. The key step after that, is to subject the graph to a simulated annealing (SA) procedure \cite{kirkpatrick1983} with a cost function dependent on bond lengths. This ensures locality, \ie interactions are kept effectively short ranged.

\subsection*{Proximity Graphs} 
\label{sec:ProximityGraphs}
	
\begin{figure*}[t]
		
		\centering
		\includegraphics[width=0.33\linewidth]{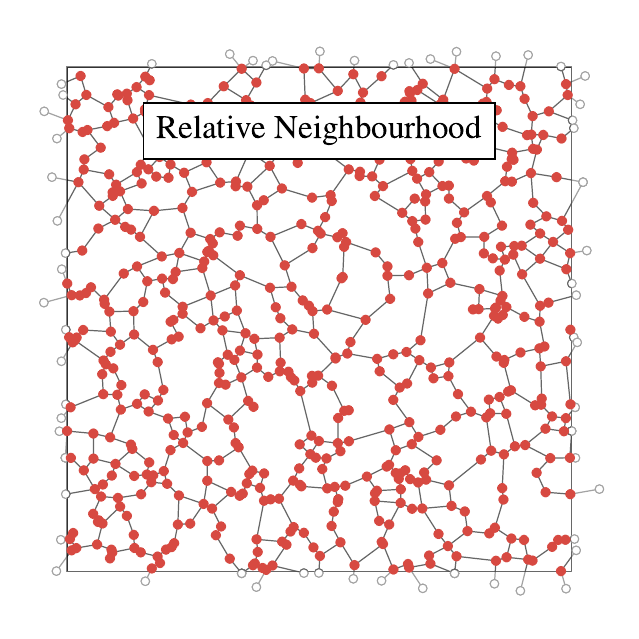}
		\includegraphics[width=0.33\linewidth]{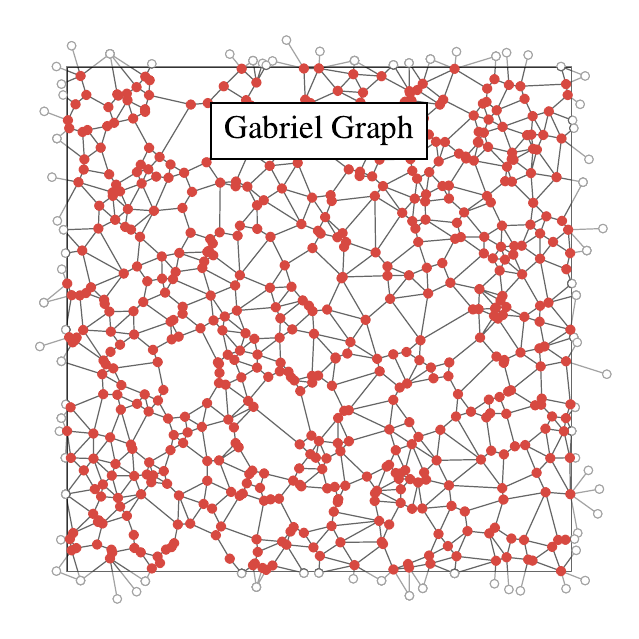}
		\includegraphics[width=0.33\linewidth]{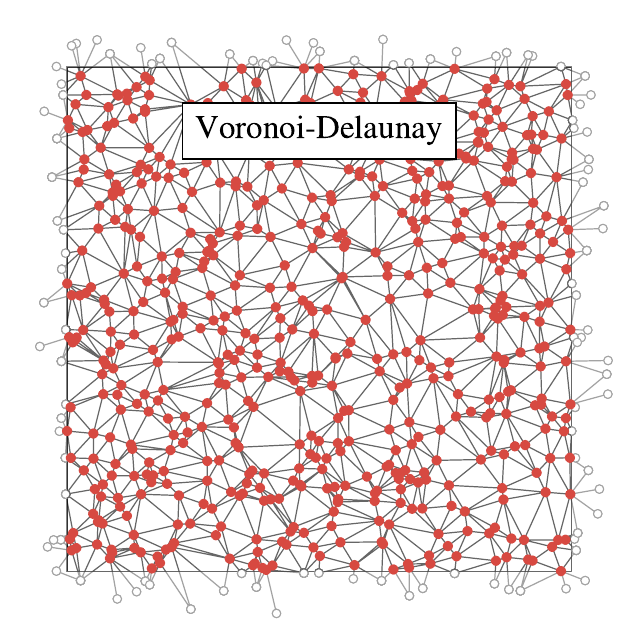} \\\vspace*{-3mm}
		\includegraphics[width=0.33\linewidth]{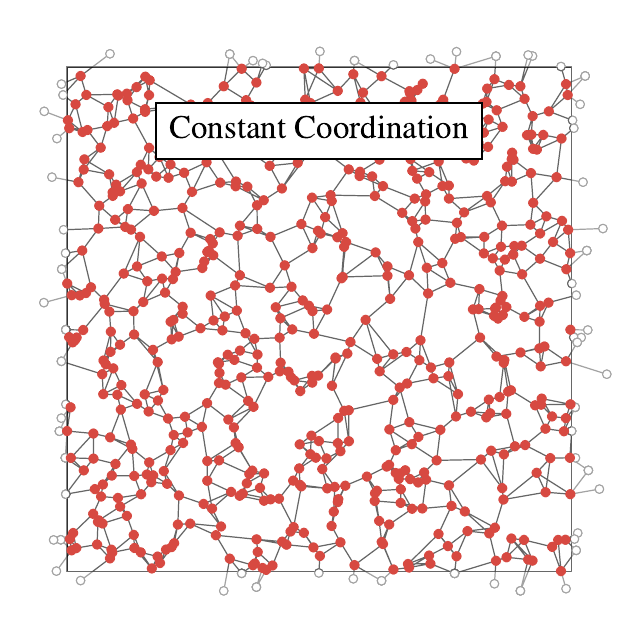}
		\includegraphics[width=0.33\linewidth]{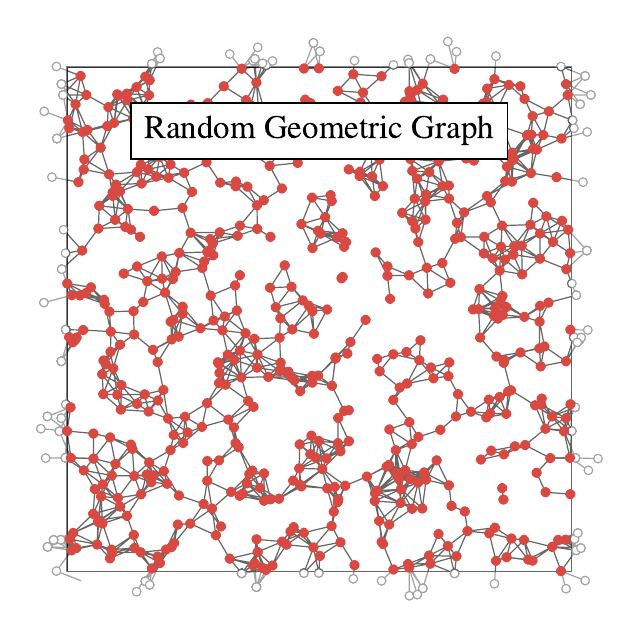}
		\includegraphics[width=0.33\linewidth]{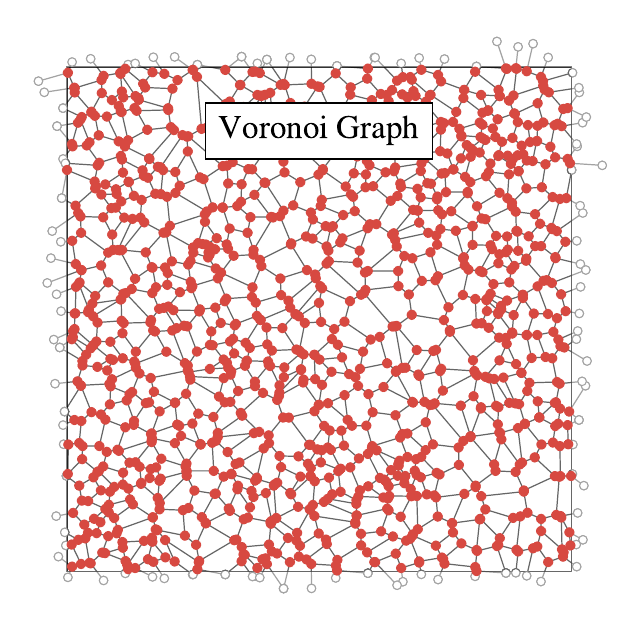} \\\vspace*{-3mm}
		\caption{Overview of different lattices considered in this paper, constructed from the same set of points.}
		\label{fig:lattice_overview}		
	\end{figure*}

	Graphs whose sites lie on a metric space and are connected whenever they are, according to a given criterion, sufficiently close together, are called proximity graphs~\cite{tamassia2013}. Different proximity criteria correspond to different graph constructions. We consider two such graphs, constructed from sets of points randomly distributed on a plane with periodic boundary conditions: the Gabriel Graph (GG)~\cite{gabriel1969} and the Relative Neighbourhood Graph (RNG)~\cite{toussaint1980}. In a GG, two sites $i$ and $j$ are connected whenever $d(i,j)^2 \le d(i,k)^2 + d(k,j)^2$ for any other point $k$ of the graph, where $d(i,j)$ denotes the distance between $i$ and $j$. The RNG is similarly defined by the more restrictive condition  $d(i,j) \le \max{[d(i,k), d(k,j)]}$. Both construction rules are illustrated in Fig.~\ref{fig:proximitygraphs}a. Another relevant graph is the Voronoi Graph (VG) and its dual, the Voronoi-Delaunay (VD) triangulation. The VG is a partition of the plane into cells such that for each cell corresponding to the site $i$, the points $k$ in that cell are closer to $i$ than to any other site $j$, see Fig.~\ref{fig:proximitygraphs}b. Also often used is the Random Geometric Graph (RGG), where any two nodes within a given distance are connected. Realizations of the described lattices can be seen in Fig.~\ref{fig:lattice_overview}.

\input{tikz_fig0}
	
\subsection*{Contact Process}
\label{sec:ContactProcess}

	The contact process is arguably the most prominent reaction-diffusion-type lattice model that exhibits a non-equilibrium phase transition. In the terminology of epidemics, each vertex of a lattice can either be infected (active) or healthy (inactive).
	During the simulation, in every time step an active site is randomly chosen. With probability $1-p$, it recovers spontaneously, \ie it becomes inactive. With probability $p$, the site stays active and a neighbour is randomly selected: if currently inactive, this neighbour becomes infected; if it is already infected, nothing happens. Either way, time is incremented by $\Delta t = 1/N_a$, where $N_a$ is the number of active sites before the update attempt.
	The behaviour of the system is controlled by the offspring probability~$p$. If offspring creation is weak, the dynamics is dominated by the annihilation process and eventually an absorbing state is reached where the whole lattice is inactive and the evolution terminates. If, instead, $p$ is large enough, the system steadily maintains an active cluster and is in the active phase. The transition between the active and absorbing phases happens at a critical probability $p_c$ whose precise value depends on the microscopic details of the lattice. At $p=p_c$ spatial and temporal correlation length scales, $\xi_\perp$ and $\xi_\parallel$, diverge and a scale-invariant, self-similar cluster emerges. On a regular lattice, the transition falls into the DP universality class.
	
	In this work we conduct two types of simulations. In the first, we start from a single infected site in an otherwise empty lattice~\cite{henkel2009} and measure three time-dependent quantities averaged over an ensemble of lattices: the average size $N_a$ of the evolving cluster, its mean-square radius $R^2$, and the survival probability $P$. At criticality, they follow a power-law behaviour
	\begin{align}
	N_a(t) \sim t^\theta, \qquad R^2(t) \sim t^{1/z}, \qquad P(t) \sim t^{-\delta},
	\label{eq:dynamical_scaling} 
	\end{align}
	where $\theta$, $ z $, and $ \delta $ denote universal exponents that characterize the transition. Note that $ R^2(t) $ is averaged over the surviving clusters only. The dynamical exponent $ z $ connects the temporal and spatial correlation lengths scales, $ \xi_\parallel\sim\xi_\perp^z$.
	If the CP is placed on a disordered lattice (such as regular lattices with diluted sites or bonds) the transition is determined by a new finite-randomness fixed point independent of the disorder strength~\cite{moreira1996,vojta2005}. There, the dynamics is ultraslow and the algebraic relations in Eq.~\eqref{eq:dynamical_scaling} are replaced by logarithmic scaling laws
	\begin{align}
	N_a(t) \sim \ln(t/t_0)^{\bar{\theta}}, \qquad R^2(t) \sim \ln(t/t_0)^{1/\psi}, \qquad P(t) \sim \ln(t/t_0)^{-\bar{\delta}}.
	\label{eq:activated_dynamical_scaling} 
	\end{align}
	In this so-called \emph{activated scaling} scenario, where $t_0$ denotes a non-universal timescale, the relation between temporal and spatial correlations becomes $ \ln(\xi_\parallel/t_0)\sim\xi_\perp^\psi$ rendering the dynamical exponent $ z $ formally infinite and giving rise to a new exponent $\psi$, called the tunnelling exponent. The disorder exponents $\bar{\delta}$, $ \bar{\theta} $ and $\psi$ have been determined numerically\cite{vojta2009,wada2017} and found to match those of the random transverse-field Ising model (RTFIM) universality class, which is also supported by analytical studies of the strong-disorder renormalization group~\cite[and references therein]{hoyos2008}. 
	
	Once the critical point is known, we perform quasi-stationary (QS) simulations \cite{oliveira2005}. In this approach, the run takes place on a finite lattice and starts from a configuration where all sites are active. Every time the absorbing state is reached, the system is reset to an active configuration randomly chosen from its history. Numerically, the system history consists in a sample of previously visited states, periodically updated to assure convergence. We measure the QS density, which is known to scale as
	\begin{align}
	\rho_\mathrm{QS}\sim L^{-\beta/\nu_\perp}
	\label{eq:qs_density}
	\end{align}
	directly at the critical point\cite{dickman2005}. For more details on the QS method, we refer the reader to Refs.~\citen{dickman2005,oliveira2005}.

%% file: tikz_fig0.tex
\begin{figure}
	\centering
\tikzset{mystyle/.style={draw, circle, semithick, fill=black!10!red, draw=black, text=white, font=\itshape, inner sep=0pt, minimum size=6pt}}

\tikzset{mystyle2/.style={draw, circle, semithick, fill=black!20!green, draw=black, text=white, font=\itshape, inner sep=0pt, minimum size=6pt}}

\begin{minipage}{0.29\textwidth}
	\centering
\begin{tikzpicture}[scale=0.45]

\begin{turn}{-8}
\tikzset{
	hatch distance/.store in=\hatchdistance,
	hatch distance=10pt,
	hatch thickness/.store in=\hatchthickness,
	hatch thickness=0.3pt
}

\makeatletter
\pgfdeclarepatternformonly[\hatchdistance,\hatchthickness]{northeast}
{\pgfqpoint{0pt}{0pt}}
{\pgfqpoint{\hatchdistance}{\hatchdistance}}
{\pgfpoint{\hatchdistance-1pt}{\hatchdistance-1pt}}%
{
	\pgfsetcolor{\tikz@pattern@color}
	\pgfsetlinewidth{\hatchthickness}
	\pgfpathmoveto{\pgfqpoint{0pt}{0pt}}
	\pgfpathlineto{\pgfqpoint{\hatchdistance}{\hatchdistance}}
	\pgfusepath{stroke}
}

\pgfdeclarepatternformonly[\hatchdistance,\hatchthickness]{northwest}
{\pgfqpoint{0pt}{0pt}}
{\pgfqpoint{\hatchdistance}{\hatchdistance}}
{\pgfpoint{\hatchdistance-1pt}{\hatchdistance-1pt}}%
{
	\pgfsetcolor{\tikz@pattern@color}
	\pgfsetlinewidth{\hatchthickness}
	\pgfpathmoveto{\pgfqpoint{\hatchdistance}{0pt}}
	\pgfpathlineto{\pgfqpoint{0pt}{\hatchdistance}}
	\pgfusepath{stroke}
}

\pgfdeclarepatternformonly[\hatchdistance,\hatchthickness]{horizontal}
{\pgfqpoint{0pt}{0pt}}
{\pgfqpoint{\hatchdistance}{\hatchdistance}}
{\pgfpoint{\hatchdistance-1pt}{\hatchdistance-1pt}}%
{
	\pgfsetcolor{\tikz@pattern@color}
	\pgfsetlinewidth{\hatchthickness}
	\pgfpathmoveto{\pgfqpoint{0pt}{0pt}}
	\pgfpathlineto{\pgfqpoint{\hatchdistance}{0pt}}
	\pgfusepath{stroke}
}

\pgfdeclarepatternformonly[\hatchdistance,\hatchthickness]{vertical}
{\pgfqpoint{0pt}{0pt}}
{\pgfqpoint{\hatchdistance}{\hatchdistance}}
{\pgfpoint{\hatchdistance-1pt}{\hatchdistance-1pt}}%
{
	\pgfsetcolor{\tikz@pattern@color}
	\pgfsetlinewidth{\hatchthickness}
	\pgfpathmoveto{\pgfqpoint{0pt}{0pt}}
	\pgfpathlineto{\pgfqpoint{0pt}{\hatchdistance}}
	\pgfusepath{stroke}
}
\makeatother

\draw ([shift={((0,0))}]-80:4) arc[radius=4, start angle=-80, end angle=90];
\draw ([shift={((4,0))}]100:4) arc[radius=4, start angle=100, end angle=270];
\draw (2,0) circle (2);
\draw[pattern=northeast, hatch distance=15pt, hatch thickness = 0.3pt] 
	([shift={((0,0))}]-60:4) arc[radius=4, start angle=-60, end angle=60] 
	([shift={((4,0))}]120:4) arc[radius=4, start angle=120, end angle=240];
	
\draw[pattern=northwest, hatch distance=15pt, hatch thickness = 0.3pt] 
(2,0) circle (2);

\coordinate[](P1) at (0,0) {};
\coordinate[](P2) at (4,0) {};
\node[mystyle] () at (P1) {};
\node[mystyle] () at (P2) {};
	\end{turn}

	\draw (-1, -4.1) node[anchor=center] {a)};

\end{tikzpicture}
\end{minipage}
\begin{minipage}{0.29\textwidth}
	\centering
\begin{tikzpicture}[scale=0.45]
	\draw (-6,-5) node[anchor=center] {b)};
\coordinate[](O) at (0,0) {};
\coordinate[](A2) at ($(O)+(10:3)$);
\coordinate[](A3) at ($(O)+(45:3)$);
\coordinate[](A4) at ($(O)+(100:3)$);
\coordinate(C5) at ($(O)+(160:4)$);
\coordinate[](K) at ($(O)+(220:3)$);
\coordinate(C1) at ($(O)+(-60:4)$);
\coordinate(B1) at ($(K)+(110:4)$);
\coordinate[](B2) at ($(K)+(170:3)$);
\coordinate[](B3) at ($(K)+(220:3)$);
\coordinate[](B4) at ($(K)+(-90:3)$);
\coordinate[](B5) at ($(K)+(-45:3.5)$);
\coordinate(B7) at ($(K)+(-10:2)$);

\coordinate[] (A5) at (intersection cs:first line={(O)--(C5)}, second line={(K)--(B1)});
\coordinate[] (B1) at (intersection cs:first line={(O)--(C5)}, second line={(K)--(B1)});
\coordinate[] (A1) at (intersection cs:first line={(O)--(C1)}, second line={(K)--(B7)});
\coordinate(B6) at (intersection cs:first line={(O)--(C1)}, second line={(K)--(B7)});

\begin{scope}[dashed,black]
	\draw (A1)--(A2)--(A3)--(A4)--(A5)--(B2)--(B3)--(B4)--(B5)--cycle;
	\foreach \x in {1,2,3,4,5}
	{
		\draw (O)--(A\x);
	}
	\foreach \x in {1,2,3,4,5,6}
	{
		\draw (K)--(B\x);
	}
	\draw (O)--(K);
\end{scope}


\foreach \i in {1,2,3,4,5}
{
	\coordinate(M\i) at ($(O)!0.5!(A\i)$);
}
\foreach \i in {1,2,3,4,5,6}
{
	\coordinate(X\i) at ($(K)!0.5!(B\i)$);
}
\foreach \i in {1,2,3,4,5}
{
	\coordinate(L\i) at  ($(M\i)!2cm!90:(A\i)$);
}
\coordinate (T1) at (intersection cs:first line={(M1)--(L1)}, second line={(M2)--(L2)});
\coordinate (T2) at (intersection cs:first line={(M2)--(L2)}, second line={(M3)--(L3)});
\coordinate (T3) at (intersection cs:first line={(M3)--(L3)}, second line={(M4)--(L4)});
\coordinate (T4) at (intersection cs:first line={(M4)--(L4)}, second line={(M5)--(L5)});

\foreach \i in {1,2,3,4,5,6}
{
	\coordinate(H\i) at  ($(X\i)!2cm!90:(B\i)$);
}
\coordinate (T5) at (intersection cs:first line={(M5)--(L5)}, second line={(X1)--(H1)});
\coordinate (T6) at (intersection cs:first line={(X1)--(H1)}, second line={(X2)--(H2)});
\coordinate (T7) at (intersection cs:first line={(X2)--(H2)}, second line={(X3)--(H3)});
\coordinate (T8) at (intersection cs:first line={(X3)--(H3)}, second line={(X4)--(H4)});
\coordinate (T9) at (intersection cs:first line={(X4)--(H4)}, second line={(X5)--(H5)});
\coordinate (T10)at (intersection cs:first line={(X5)--(H5)}, second line={(X6)--(H6)});
\coordinate (T11)at (intersection cs:first line={(M1)--(L1)}, second line={(X6)--(H6)});

\begin{scope}[black,semithick]
	\draw(T1)--(T2)--(T3)--(T4)--(T5)--(T6)--(T7)--(T8)--(T9)--(T10)--(T11)--cycle;
	\draw(T5)--(T11);
	\coordinate (E1) at ($(A1)!0.5!(A2)$);
	\coordinate (E2) at ($(A2)!0.5!(A3)$);
	\coordinate (E3) at ($(A3)!0.5!(A4)$);
	\coordinate (E4) at ($(A4)!0.5!(A5)$);
	\coordinate (E5) at ($(A5)!0.5!(B2)$);
	\coordinate (E6) at ($(B2)!0.5!(B3)$);
	\coordinate (E7) at ($(B3)!0.5!(B4)$);
	\coordinate (E8) at ($(B4)!0.5!(B5)$);
	\coordinate (E9) at ($(B5)!0.5!(B6)$);
	\draw(T1)--($(T1)!2!(E1)$);
	\draw(T2)--($(T2)!2!(E2)$);
	\draw(T3)--($(T3)!2!(E3)$);
	\draw(T4)--($(T4)!2!(E4)$);
	\draw(T6)--($(T6)!2!(E5)$);
	\draw(T7)--($(T7)!2!(E6)$);
	\draw(T8)--($(T8)!2!(E7)$);
	\draw(T9)--($(T9)!2!(E8)$);
	\draw(T10)--($(T10)!2!(E9)$);
\end{scope}

\node[mystyle] () at (A1) {};
\node[mystyle] () at (A2) {};
\node[mystyle] () at (A3) {};
\node[mystyle] () at (A4) {};
\node[mystyle] () at (A5) {};
\node[mystyle] () at (B1) {};
\node[mystyle] () at (B2) {};
\node[mystyle] () at (B3) {};
\node[mystyle] () at (B4) {};
\node[mystyle] () at (B5) {};
\node[mystyle] () at (K) {};
\node[mystyle] () at (O) {};

\node[mystyle2] () at (T1) {};
\node[mystyle2] () at (T2) {};
\node[mystyle2] () at (T3) {};
\node[mystyle2] () at (T4) {};
\node[mystyle2] () at (T5) {};
\node[mystyle2] () at (T6) {};
\node[mystyle2] () at (T7) {};
\node[mystyle2] () at (T8) {};
\node[mystyle2] () at (T9) {};
\node[mystyle2] () at (T10) {};
\node[mystyle2] () at (T11) {};

\end{tikzpicture}
\end{minipage}
	\caption{(a) GG (RNG) construction: the smallest circle (lune) defined by two connected sites, indicated by the cross- (single-) hatching, should contain no other sites. (b) Illustration of the VG (green) and VD (red) lattices.}
\label{fig:proximitygraphs}		
\end{figure}

%% file: results.tex
\section*{Results}
\label{sec:Results}
	
	\begin{figure*}[t]
		\centering
		\includegraphics[width=\linewidth]{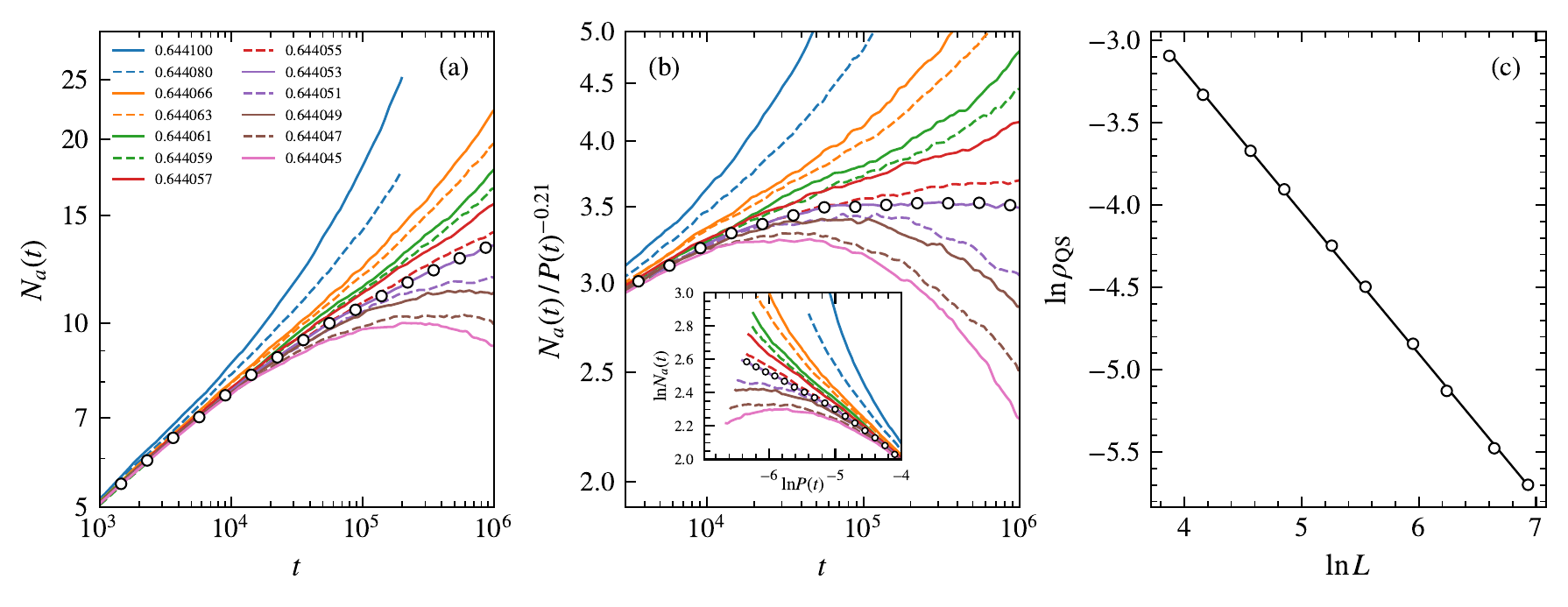}
		\caption{Numerical results for the CC4. (a) Average cluster size as a function of time for spreading runs. The markers highlight the curve corresponding to the estimated critical point. (b) Verification of the exponent $\theta/\delta$. (c) Quasi-stationary density against the linear system size $L$. The solid line represents a linear fit to the data points, with slope $\beta/\nu$ (see Tab.~\ref{tab:exponents}). 
		}
		\label{fig:seed_results_cc}		
	\end{figure*}
	
	We perform simulations of the contact process on the CC lattice with four neighbours at each node (CC4), as well as on the GG and RNG. For the seed simulations we use lattices of linear size $L=12\,000$ with periodic boundary conditions. This large size allows simulation times at least up to $ T=10^6 $, while avoiding any finite-size effects.
	Even larger lattices could not be afforded due to the huge memory requirements. In total we used up to 600 independent disorder realizations, with 10\,000 spreading runs per probability on each realization.
	In the following we present the analysis for the CC4. The results for the RNG and GG are completely analogous and can be found in the Supplementary Material.
	
	\begin{table}[b]
		\centering
		\begin{tabular}{cccccc}
			& $ \theta/\delta $ & $ \theta\psi $ & $ \delta\psi $  & $ \beta/\nu_\perp$ & $ p_c $\vspace*{1mm}\\
			\hline \\\vspace*{1mm}
			CC                      & $0.21\substack{+0.03 \\ -0.08}$  & 0.10(3)   & 0.53(3)   & 0.85(1)[1] & 0.644053(2) \\\vspace*{1mm}
			RNG                     & $0.17\substack{+0.05 \\ -0.04}$  & 0.10(2)   & 0.54(2)   & 0.88(1)[1] & 0.672389(3) \\\vspace*{1mm}
			GG                      & $0.23\substack{+0.06 \\ -0.10}$  & 0.12(4)   & 0.52(4)   & 0.84(1)[1] & 0.631767(3)\vspace*{1mm}\\
			\hline \\ \vspace*{1mm}
			DP reference~\cite{dickman1999}                 & 0.507(1)  & 0.4051(7) & 0.799(2) & 0.799(2)   &  \\
			RTFIM reference~\cite{wada2017} & 0.075(5) & 0.078(4)  & 1.034(23) & 0.964(2)   & \\ 
			
		\end{tabular} 
		\caption{Critical exponent results for the lattice models considered in this work. Errors stem from the uncertainty of the critical point and fluctuations of the individual data points. The latter are indicated in square brackets whenever comparably large to the former. The last two lines show the corresponding exponent combinations for the DP universality class~\cite{dickman1999} and for the random transverse-field Ising model~\cite{wada2017} as reference values. Note that in the case of DP universality, $\delta z = \beta/\nu_{\perp}$. } \label{tab:exponents}
	\end{table}

	The first step in the analysis is to locate the critical point. According to the criterion of Dickman \cite{moreira1996}, we search for the smallest $ p $ that results in a curve that asymptotically does not decay. In Fig.~\ref{fig:seed_results_cc} we therefore evaluate $ N_a(t) $ for several probabilities. The curves corresponding to the smallest values decay quickly, while those for the largest $ p $ bend away from the critical region, providing us with a rough estimate of the critical value. In order to obtain a more precise estimate of $ p_c $ we plot $ N_a(t) $ against $ P(t) $ (Fig.~\ref{fig:seed_results_cc}b, inset) and search for the curve best described by the power law of Eq.~\eqref{eq:activated_dynamical_scaling}, \ie~the one which is straight in the long-time limit. We find $ p_c=0.644053(2) $, where the error is estimated from the nearby curves that show noticeable deviations from a power law. Fitting Eq.~\eqref{eq:activated_dynamical_scaling} to the asymptotic region (which spans about one order of magnitude in $ P(t)$) we obtain the exponent combination $ \theta/\delta = 0.21\substack{+0.03 \\ -0.08}$ where the error stems from the uncertainty of the critical point. We verified this estimate by also plotting the ratio $N(t)/P(t)^{-\theta/\delta}$, and found the expected results (asymptotically constant behavior) to be fulfilled best at $ \theta/\delta = 0.21$ (Fig.~\ref{fig:seed_results_cc}b), confirming the former estimate. 
	A similar analysis of $ N_a $ vs $ R^2 $ and $ P $ vs $ R^2 $, yields the exponent combinations $ \theta\psi = 0.10(3)$ and $ \delta\psi=0.53(3) $. Due to the very late onset of the non-algebraic behaviour we can estimate only exponent ratios rather than the values of $\theta$, $\delta$ and $\psi$. In principle, those could be computed by fitting Eqs.~\eqref{eq:activated_dynamical_scaling}  to the respective curves under the constraint of $\ln(t_0)$ being the same for the three observables. However, as stable fits would require much longer simulations times and therefore enormous lattice sizes, this lies beyond present computational capabilities.
	
	For the QS simulations we use lattice sizes up to $L=1024$ and simulate the system for $2\cdot 10^8$ time steps. For the largest lattices we use at least 140 disorder realizations, for the smaller ones up to 1000 realizations. The quasi-stationary density is shown in Fig.~\ref{fig:seed_results_cc}c, revealing the expected straight behaviour of Eq.~\eqref{eq:qs_density}. A linear fit yields the estimate $ \beta/\nu_\perp = 0.85(1)[1] $, where the first error stems from the uncertainty of the critical point whereas the second one is due to the statistical fluctuations of the individual data points in the fit ($ \chi^2/\mathrm{d.o.f.}=1.64 $). The exponent estimates, also for the RNG and GG, are listed in Tab.~\ref{tab:exponents}. 

%% file: discussion.tex
\section*{Discussion}
\label{sec:Discussion}
	
	The critical exponent combinations we obtain for the topologically disordered lattices, compared to those of the clean DP universality class (Tab.~\ref{tab:exponents}), make it evident that, despite the relatively large error bars, it can be ruled out that the phase transitions belong to this class. This is already indicated by the behaviour of the dynamic observables, which show no straight lines in the double logarithmic plots (Fig.~\ref{fig:seed_results_cc}, curves in the vicinity of the critical point). Furthermore, the exponents of all three models do not match those of the RTFIM either, which rules out this scenario as well. Strikingly, however, the exponents for the CC, RNG, and GG simulations coincide within their error bars, suggesting that the contact process on those three lattices is determined by a yet undiscovered disorder fixed point and the three models belong to the same (new) universality class. Whether or not that is the case remains open, especially since topological disorder has so far not been within the reach of analytical RG methods. From the numerical standpoint, more precise estimates of the exponent combinations can reveal whether they remain compatible with each other within smaller uncertainties or if there exists a weak dependence on the disorder strength that our simulations are not able to resolve. Given the already considerable computational effort demanded for obtaining the present results, these questions also remain open.
	
	Irrespective of the precise classification of the phase transition in our models, the fact that they do not belong to the DP universality class means that disorder is relevant in all three cases. But what makes those types of disorder relevant? Barghathi and Vojta\cite{barghathi2014} generalized Harris' relevance criterion\cite{harris1974}, arguing that coordination-number fluctuations that decay slowly enough disrupt the collective behaviour at criticality, i.e. change the nature of the phase transition. However, not all fluctuations are equal: As long as the edges remain local, slowly-decaying fluctuations caused by a surplus of edges will not deter the emergence of collective behaviour, as we have recently shown\cite{schrauth2018b}. Thus, the relevant inhomogeneities seem to be those that impair the propagation of information through the system. Therefore, the key concept appears to be some form of \emph{poor connectivity} of the lattice and how fast it decays.

\subsection*{Dual Tessellation}
\label{sec:GeneralizedDualLattice}

	In order to characterize this poor connectivity we turn our attention to \emph{holes} in the lattice. A hole is understood as a region where nodes metrically close to each other are separated by large edge distances. Pictorially, holes are edgeless regions in the lattice where, were the edges between the nodes distributed more uniformly in space (such as in a triangulation), edges would be expected. In planar graphs, holes trivially correspond to faces, and a large hole is a face surrounded by a large number of nodes. The VD, for instance, has only triangular faces and, therefore, no proper holes, while GGs have moderately-sized holes and RNGs relatively large holes, see Fig.~\ref{fig:lattice_overview}. For planar graphs a convenient way to analyse the size of these holes (faces) is through the so-called \emph{dual graphs}, where each site corresponds to a given face of the original lattice and bonds are placed between those sites whose corresponding faces share an edge~\cite{harary1969}. Thus, dual lattice sites with high coordination numbers correspond to large holes in the original graph. In this way, following Barghathi and Vojta~\cite{barghathi2014}, by considering the coordination-number fluctuations on their dual lattices, we can investigate the effect of hole-size fluctuations and their spatial decay on the universality of processes on the original lattices.
	
	When considering \emph{non-planar} graphs, such as the CC lattice, the concept of faces, and therefore dual graphs, is ill-defined and we need a new tool. Given that a face is delimited by the smallest path around it, we propose a generalization of face, which is a \emph{polygon} $ \mathcal{P}$, defined by a closed path, whose \emph{weight} $w$ corresponds to the size of the face. These polygons can overlap with each other and, together, constitute a \emph{dual tessellation} whose end-product is a function $W(x)$ attributing a weight to each point of space. A thorough description of this construction is presented in the Appendix~\ref{sec:DualTessellation}.
	
\begin{figure}
		\centering
		\includegraphics[width=1.0\linewidth]{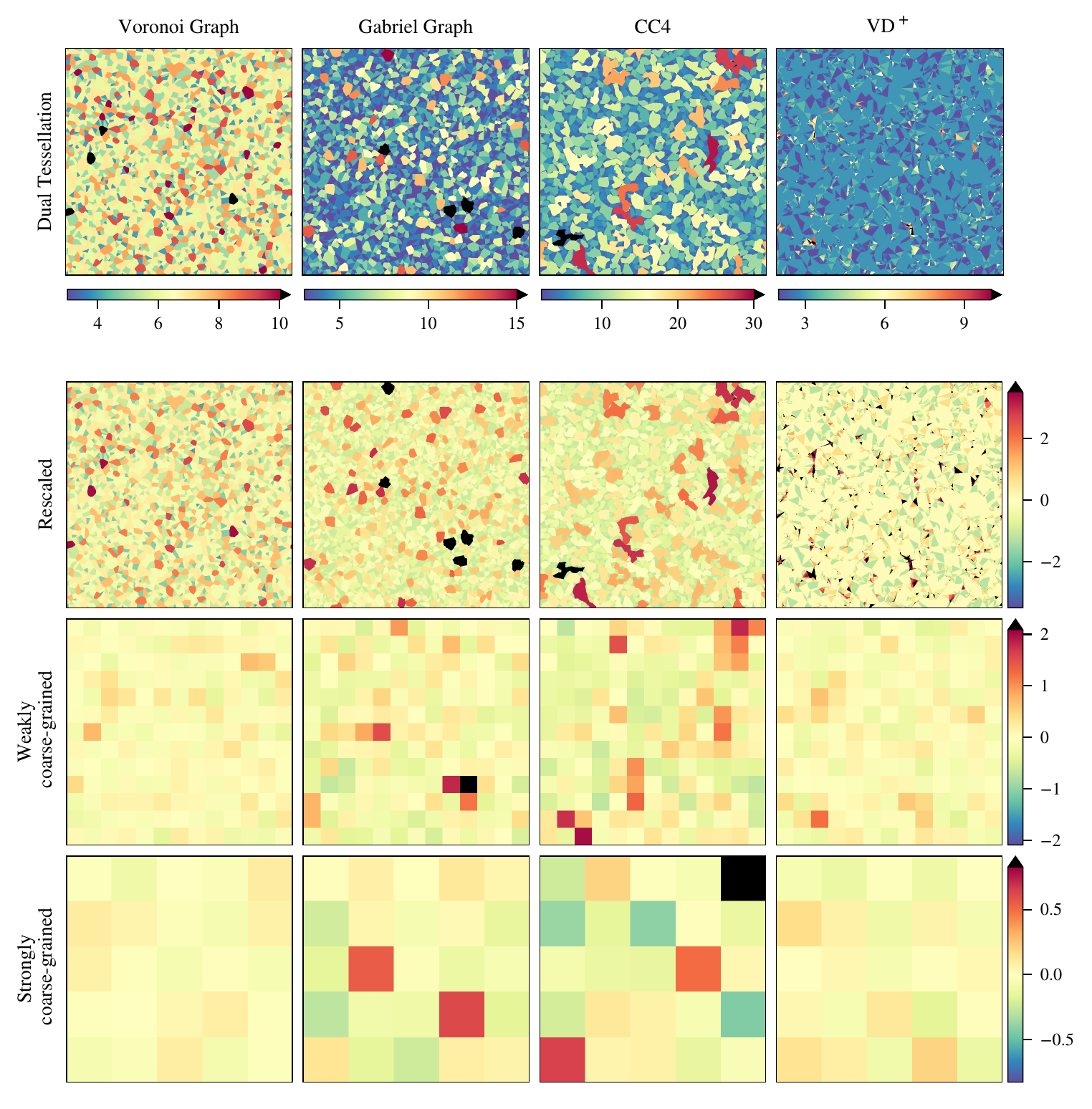}
		\caption{Dual tessellation samples for the lattices VG, GG, CC4, and VD$^+$. In the second row, the dual tessellation weights are rescaled
		as described in the text. Hence, the colours denote the fluctuations of the weights around their respective spatial averages. Except for the top row, the colour ranges are the same in every row. The two bottom rows show coarse graining steps.}
		\label{fig:dual_tessellation_cg}		
	\end{figure}

	We calculate the dual tessellation for the VG, GG, CC4 and VD$^+$, shown in the top row of Fig.~\ref{fig:dual_tessellation_cg}. In planar graphs, (ordinary) faces present no overlaps and the weights are simply the coordination numbers of the corresponding sites of the dual lattice. Hence, the VG's dual tessellation shows the coordination number fluctuations of the corresponding VD lattice, its dual. As can be seen, the weight amplitudes vary considerably between the different lattices considered. For instance, VG cells with $W>10$ are rare, whereas the CC features a few large holes with $W\approx30$. In order to make the weight fluctuations comparable, we rescale the dual tessellations according to 
	$ W_\mathrm{rescaled} = (W-\mu_{W})/\sigma_W $, where $\mu_W$ and $\sigma_W$ represent the weight mean and standard deviation.
	
	The rescaled dual tessellations for the four lattices displayed in the second row of Fig.~\ref{fig:dual_tessellation_cg} are quite distinct: the VG is characterized by regions of very high and very low weights found close together, giving it a relatively homogeneous, well-mixed appearance -- a visual consequence of the coordination number anti-correlation present in the VD lattice~\cite{barghathi2014}; the GG's dual tessellation, on the other hand, appears less homogeneous, with high-weight regions standing out more clearly, as a consequence of the lack of such anti-correlations; even less homogeneous is the CC4, whose dual tessellation displays large swaths of high-$W$ regions against a relatively featureless background; and the VD$^+$ is marked by randomly distributed peaks of very high weights. More decisive still than the magnitude of the fluctuations, is how fast they decay under a block coarse-graining analysis~\cite{barghathi2014}, i.e. how fast homogeneity increases as the weights are averaged over increasingly larger square blocks of the spacial domain. The lower two rows of Fig.~\ref{fig:dual_tessellation_cg} show two steps of this analysis. In the case of the VG, significant homogeneity sets in already after a weak coarse-graining step, due to the inherent anti-correlations mentioned above. This fast decay of the fluctuations indicates that the disorder is less relevant. In contrast, the GG has no such anti-correlations~\cite{schrauth2018a} and regions of high weight survive large coarse-graining steps, keeping the tessellation less homogeneous through the procedure. For the CC lattice -- an important motivation for the introduction of the generalized dual tessellation -- we find that the fluctuations are indeed as pronounced, and decrease equally slowly as those of the GG. Finally, we see that the prominent peaks of the dual tessellation of the VD$^+$ quickly disappear under coarse-graining, leaving only fluctuations that decay about as fast as those of the VG.
	
	Just as the behaviour of coordination number fluctuations under coarse graining of the original (\ie non-dual) lattice is used to establish the HBV criterion, we propose to apply the same reasoning to the \emph{dual} tessellation. This means taking a slow decay in the dual tessellation weights to imply a stronger relevance of disorder in the original lattice. Hence, the slow decay of fluctuations in the CC4 dual tessellation indicates that the disorder is as relevant as for the GG, agreeing with the results of our simulations (Tab.~\ref{tab:exponents}), where both lattices show very similar behaviour.
	Therefore, we argue that the relevance of the disorder present in the contact process on the CC4 (see Sec.~\ref{sec:Results}) is reflected by the persistent fluctuations in its dual tessellation. 
	Conversely, the fast decay of dual-tessellation fluctuations for the VD$^+$ would indicate its disorder to be less relevant. Hence, it might successfully explain the universal character of the CP, which could not be explained by a coordination number analysis performed directly on the VD$^+$ lattice~\cite{schrauth2018b}.

	\vspace*{-0.3em}
\subsection*{Elastic Relaxation}
	
\begin{figure*}
		\newcommand{\sfsize}{0.30}
		\centering
		\includegraphics[width=\sfsize\linewidth]{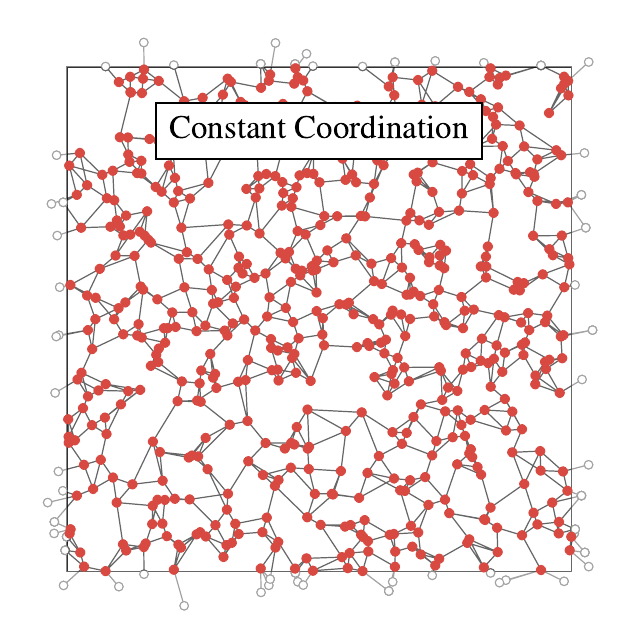}
		\includegraphics[width=\sfsize\linewidth]{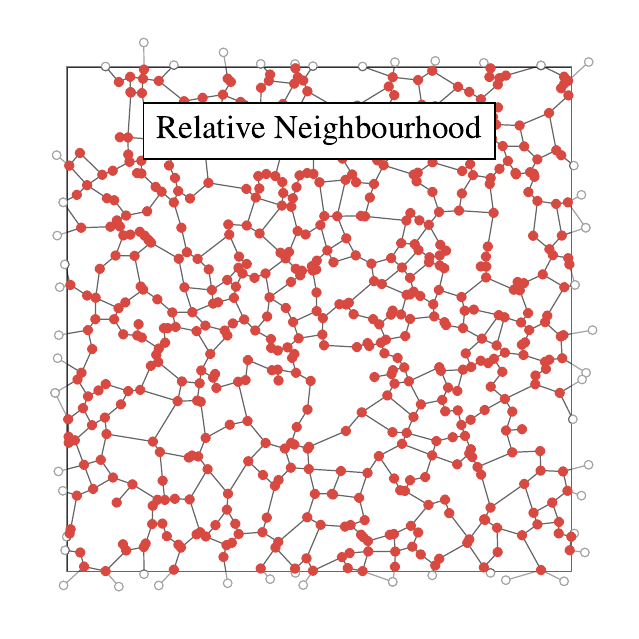}
		\includegraphics[width=\sfsize\linewidth]{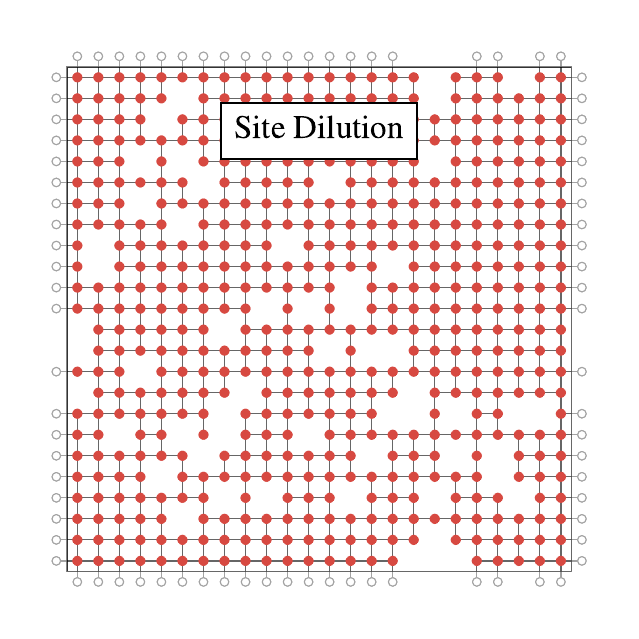} \vspace*{-3mm}
		
		\includegraphics[width=\sfsize\linewidth]{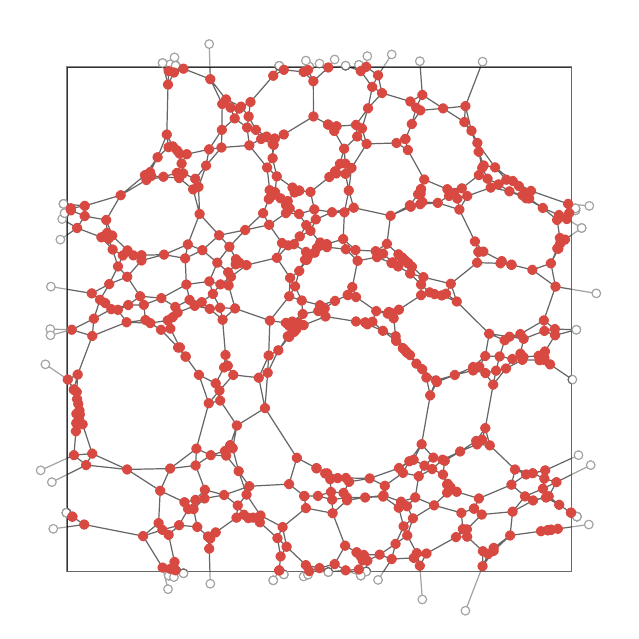}
		\includegraphics[width=\sfsize\linewidth]{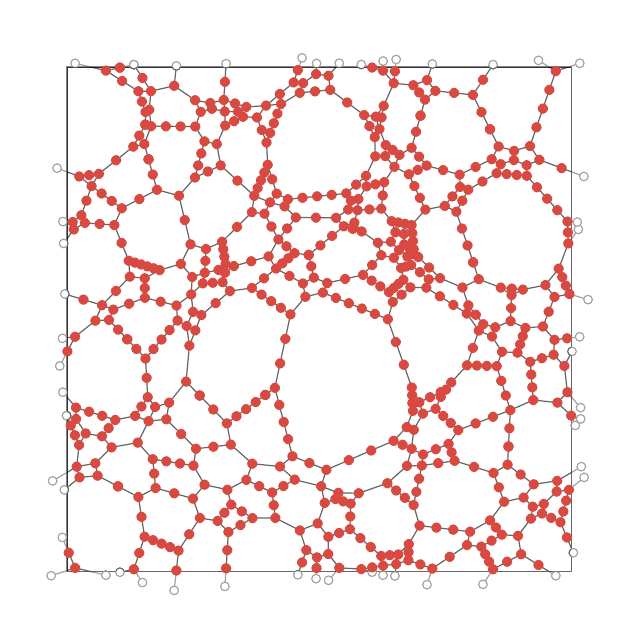}
		\includegraphics[width=\sfsize\linewidth]{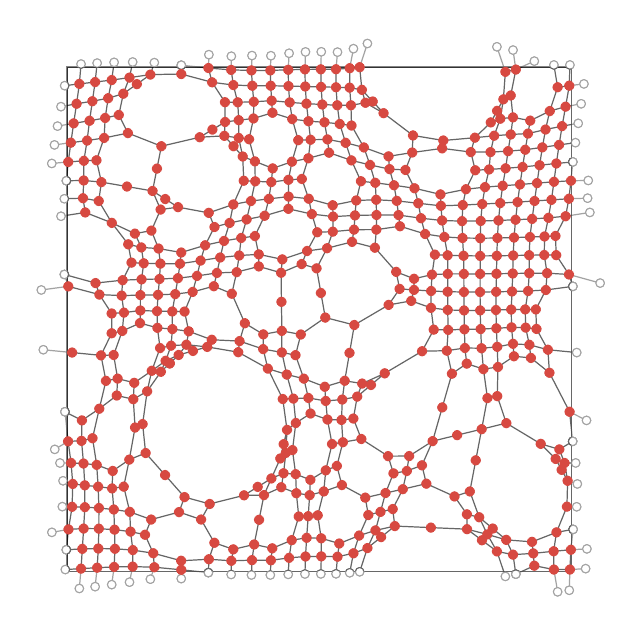}\\\hspace*{8mm}
		
		\includegraphics[width=\sfsize\linewidth]{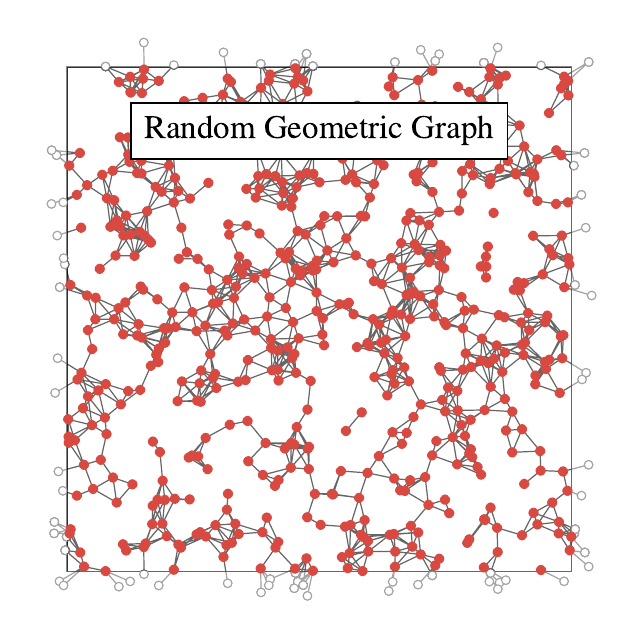}	
		\includegraphics[width=\sfsize\linewidth]{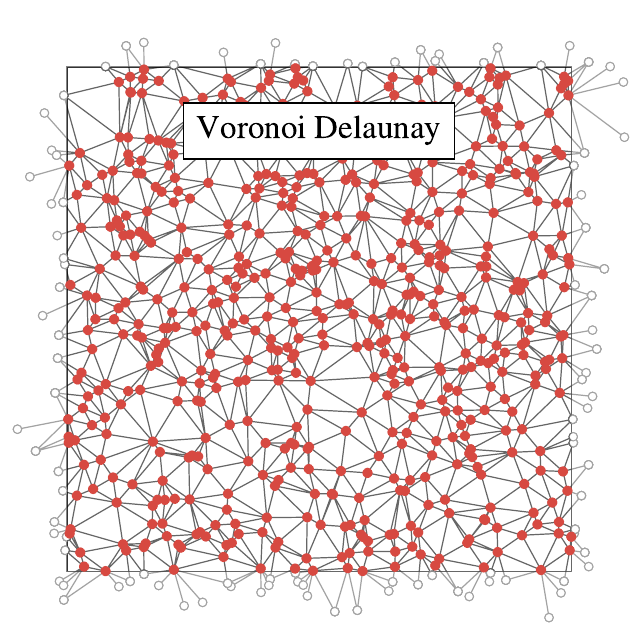}	
		\includegraphics[width=\sfsize\linewidth]{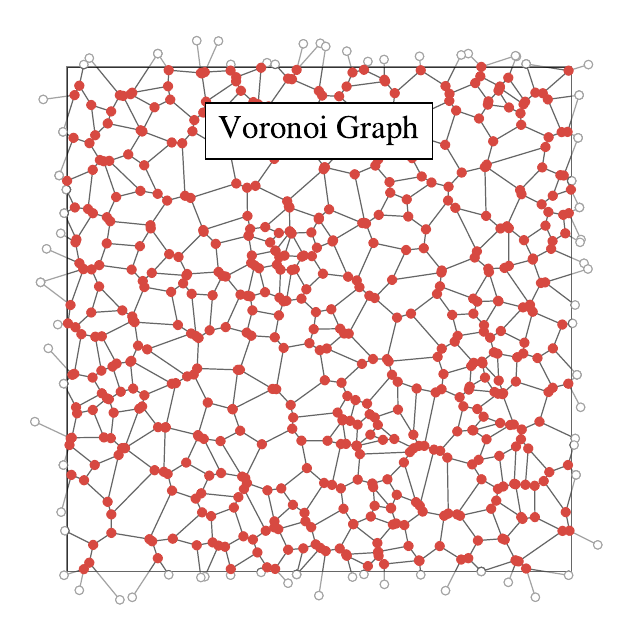}	\vspace*{-3mm}
		
		\includegraphics[width=\sfsize\linewidth]{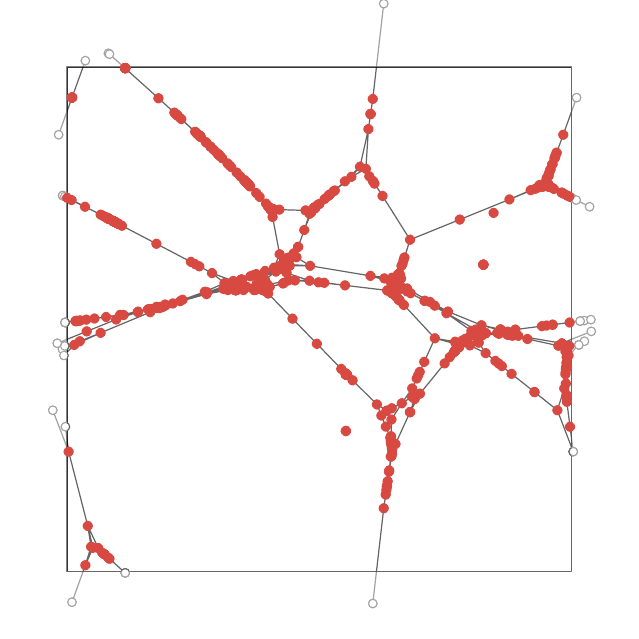}	\includegraphics[width=\sfsize\linewidth]{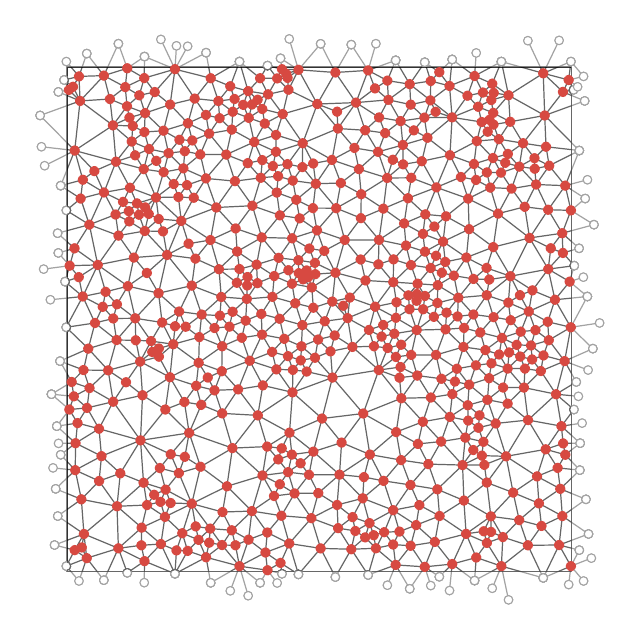}	\includegraphics[width=\sfsize\linewidth]{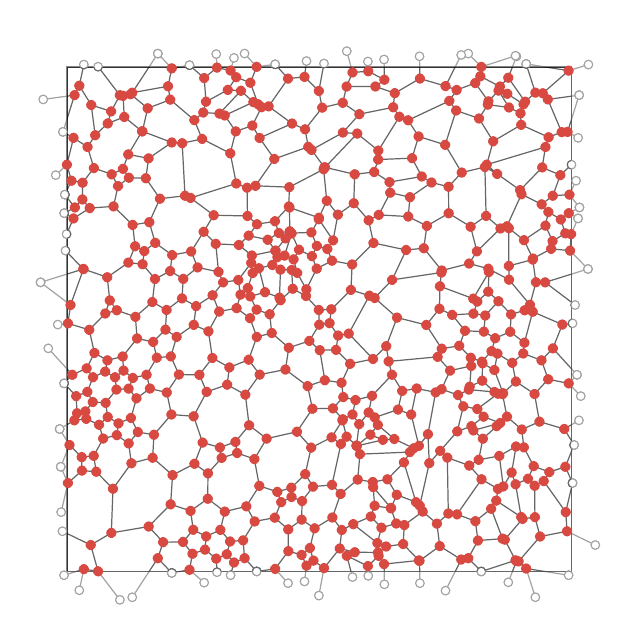}
		\caption{Random lattices before and after elastic relaxation. Pulled by tensions attributed to the bonds, sites tend to agglutinate, accentuating any holes present. CC, RNG, RGG and VD are generated from the same set of points.}
		\label{fig:before_and_after}		
	\end{figure*}

	In order to make the spatial distribution of nodes, initially random, reflect the topology of the lattice, we perform a relaxation process with respect to tensions imposed along bonds. We let one site at a time move, under a decreasing heat bath, according to the total force the site is subjected to when we take its bonds to be stretched Hookean springs. Lattice sites are therefore attracted to sites they are connected to, and bond-poor regions tend to expand -- highlighting the poor connectivity we believe to characterize relevant topological disorder. In this way, the topology of the lattice becomes accessible to \emph{geometrical} tools, besides the established tools from graph theory. This approach also greatly facilitates visual inspection in 2D lattices, since any holes present become more pronounced, as can be seen in Fig.~\ref{fig:before_and_after}, and it can trivially be generalized to higher dimensions.
	
	Sites in periodic and quasiperiodic structures find themselves in equilibrium or near equilibrium with respect to bond tension and wiggle about slightly due to the heat bath, but otherwise remain unchanged under elastic relaxation. As soon as dilution is introduced into the lattice, however (e.g. by randomly eliminating a fraction of its sites), the picture changes: even at low dilution rates, the lattice's connectivity is affected and relaxation makes a number of significant holes evident. Similar and even more numerous holes can be seen in the relaxed versions of the proximity graphs, such as the RNG. The CC lattice also presents significant holes and a remarkable resemblance to the RNG after relaxation. Among the considered lattices, the most dramatic change brought about by elastic relaxation is found in the RGG, which is reduced to threads by the procedure. Triangulations of random clouds, on the other hand, contain maximal sets of non-crossing bonds and, akin to periodic tessellations, are little affected by relaxation, as can be seen in the VD triangulation. The introduction of a number of additional random local bonds to a VD triangulation (VD$^+$, not shown) promotes a moderate degree of agglutination, but does not otherwise change the scenario. Also the VG remains mostly unaffected by the relaxation.
	
	The results described above show that the elastic relaxation procedure provides a tool which clearly distinguishes, to the unaided eye, between lattices with and without significant holes. Every lattice belonging to the latter group (no holes: VD, VD$^+$, and VG) is found to belong to the DP universality class, while those with significant holes do not. These results provide strong, if qualitative evidence for our claim that the decisive factor for the relevance of topological disorder is how strongly the disorder affects the connectivity properties of a lattice.
	
%% file: conclusion.tex
\section*{Conclusion}
\label{sec:Conclusion}

In this work we performed large-scale numerical simulations of the contact process on several spatially disordered lattices in order to investigate the relevance of topological disorder on the DP phase transition. Of particular interest is the constant coordination lattice (CC), where each site is connected to a fixed number $q$ of local neighbours. In Ref.~\citen{schrauth2018a} two of the authors found significant deviations of the 2D Ising exponents from their clean universal values on CC lattices with $q=4$, $6$, and $10$. However, the Ising model in two dimensions, with an exponent $d\nu=2$, sits precisely on the borderline with respect to the Harris criterion -- and it remains unclear whether the CC disorder just represents a marginally relevant type of perturbation for that model, or instead the clean universal exponents can be recovered by longer simulations and accounting for strong logarithmic scaling corrections in the analysis. In contrast, the DP universality class in 2D is not marginal in terms of the Harris criterion, allowing for a clearer distinction. 

Our simulations reveal continuous absorbing-state phase transitions for RNG, GG, and CC lattices. However, the critical exponents at the transition are clearly not compatible with the clean DP reference values, strongly indicating that they do not belong to this universality class. Instead, the behaviour observed in the seed simulations suggests that some form of activated scaling is present, similar to the case of a diluted regular lattice, though noticeably weaker. Our results also rule out the random transverse-field Ising class to which belongs the CP under conventional, i.e. uncorrelated disorder, such as bond or site dilution. Interestingly, the exponents in the activated scaling scenario for the three models turn out to be compatible with each other within their errors, suggesting the possibility of a common, new universality class.

The fact that topological disorder in the RNG and GG represents a relevant perturbation to the absorbing state phase transition is not unexpected, as neither lattice fulfills the HBV criterion~\cite{barghathi2014,schrauth2018a}.
However, also for the CC lattice the clean DP universality class is clearly violated.
Since the CC lattice by construction has no coordination number fluctuations, the departure from the clean universal behaviour must originate from an implicit connectivity disorder not covered by the HBV criterion, as it only considers explicit coordination number disorder. Hence, a more general criterion is needed to describe the stability of continuous phase transitions against quenched spatial disorder.

Based on the assumption that poor connectivity may lead to relevant perturbations, we devise two tools: the dual tessellation and the elastic relaxation. The dual tessellation, closely related to the concept of dual graphs, uses cycles to reveal large topological holes in the lattice, which can then be subjected to a qualitative coarse-graining analysis. We find that for the CC4 the disorder fluctuations decay as slowly as for the GG, possibly explaining the striking similarity of results in the CP simulations for those two lattices. The dual tessellation fluctuations of the VG and VD$^+$, however, decay noticeable faster under the coarse graining, which in turn might explain why the clean DP phase transition on those lattices is retained~\cite{paper-vg,schrauth2018a}. In the elastic relaxation, we consider the lattices subjected to mechanical tension along their bonds and allow the sites to move according to the resulting forces. This leads poorly connected regions to expand, clearly revealing graphs that are loosely connected, in contrast to tightly connected graphs, such as VD, which are stable against the relaxation. As expected, poor connectivity corresponds to the CP process leaving the clean DP universality class.

These two geometric approaches are able to qualitatively distinguish which perturbations are relevant against the two-dimensional CP phase transition. In this way they can be a road sign towards a more general relevance criterion, in that they suggest that a measure of the degree of connectivity provides an additional property required to generalize existing relevance criteria.

%% file: appendix.tex
\appendix
\section{Dual Tessellation}
\label{sec:DualTessellation}

	In this section we introduce a generalization of the concept of duality to non-planar graphs, such as the CC lattice and the RGG (see Fig.~\ref{fig:lattice_overview}), for which dual graphs are not defined. Based on the fact that a face is delimited by the smallest path around it, we take as our starting point \emph{cycles $\mathcal{C}$}, which reduce, in graphs with no bond crossings, to usual faces. 
	The algorithm for finding the cycles $ \mathcal{C} $ works as follows: Consider a walker which starts from a given site, say $i$, in the lattice and travels along a given bond, ending at site $j$; from $j$ it continues along the rightmost (alternatively leftmost) bond with respect to the incoming bond, \ie it chooses the next site $k$ such that the angle between $i-j$ and $j-k$ in clockwise (anti-clockwise) direction is smallest. This rule is iterated until the walker closes a cycle, \ie returns to $i$, which is guaranteed by the finiteness of the lattice. The cycle obtained as just described is what we call a generalized face. A schematic visualization is shown in Fig.~\ref{fig:loop_algorithm}a, where a left-turning walker starting from A to B finds the cycle ABCDA.
	
	In this procedure the walker may close an intermediate loop before ultimately returning to its origin. In order to avoid inflating the cycle length and over-counting these loops (already found for different initial conditions) those intermediate, or \emph{stray} loops are pruned from the cycle. An example is shown in Fig.\ref{fig:loop_algorithm}b: A left-turning walker starting from $E$ towards $D$ travels along the path
	\begin{equation}
	\contraction{ED}{C}{FG}{C}
	\contraction{EDCFGC}{B}{A}{B}
	\text{EDCFGCBABE} = \text{EDCBE},
	\end{equation}
	where the contraction symbols mark the pruned intermediate loops, namely the triangle CFGC and the leaf A. Note that a right-turning walker starting from E towards D returns the quadrilateral path EDCBE directly. 
	In less trivial graphs the pruning process may be non-unique, in which case all possible pruned cycles are collected. An example can be seen in Fig.~\ref{fig:loop_algorithm}d for a left-turning walker starting from F to E, where three different, overlapping intermediate loops are present, resulting in three possible pruned cycles:
	\begin{equation}
	\contraction{\mathrm{F}}{\mathrm{E}}{\mathrm{DCBAD}}{E}
	\contraction[2ex]{\mathrm{FEDC}}{\mathrm{B}}{\mathrm{ADEGHIJDA}}{B}
	\bcontraction{\mathrm{FE}}{\mathrm{D}}{\mathrm{BADEGHIJx}}{D}
	\mathrm{FEDCBADEGHIJDABF} = 
	\begin{cases}
	\mathrm{FEGHIJDABF} \\ \mathrm{FEDABF} \\\mathrm{FEDCBF}.
	\end{cases}
	\label{eq:non-unique}
	\end{equation}
	Furthermore, closed paths are invariant under cyclic permutations, thus, we define equivalence classes, such as
	\begin{equation}
	[\mathcal{C}_{ABCD}] \equiv \{x\in\Omega\,\, |\,\, \mathcal{C}_{ABCD} \sim x\} = \{\mathcal{C}_{ABCD},\mathcal{C}_{BCDA},\mathcal{C}_{CDAB},\mathcal{C}_{DABC}\},
	\label{eq:equivalence_class_planar}
	\end{equation}
	where $\Omega$ denotes the set of all cycles found in the graph by the algorithm. In the planar case the number of permutations (\ie the cardinality or \emph{multiplicity} of $ [\mathcal{C}_i] $) exactly coincides with the number of edges in the cycle $ \mathcal{C}_i $ for every $ i $. Every class of equivalent cycles defines a \emph{polygon} $ \mathcal{P}([\mathcal{C}]) $ that has the cycle's sites on its vertices (corners). To the polygon a \emph{weight} $w_i$ is attributed, which is given by
	\begin{align}
	w_i\equiv |[\mathcal{C}_i]| \leq |\mathcal{C}_i|,
	\label{eq:weight}
	\end{align}
	where the equality always holds for the planar case. In the above example, the polygon $ \mathcal{P}_{ABCD} \equiv \mathcal{P}([{\mathcal{C}_{ABCD}}]) $ consists of four vertices and therefore carries the weight $ w_{ABCD} \equiv |[\mathcal{C}_{ABCD}]| = |\mathcal{C}_{ABCD}| = 4 $.

\input{tikz_fig1}

	For non-planar graphs, not necessarily every possible permutation is found by the walker, hence in general $|[\mathcal{C}_i]| \leq |\mathcal{C}_i|$. Furthermore, when the pruning process is non-unique, we attribute fractional weights to the pruned paths. In Eq.~\eqref{eq:non-unique} each of the three pruned paths contributes to a different set of equivalent cycles with weight 1/3. 
	Consider the following example, where the overtext shows how often each cycle was found by the algorithm
	\begin{equation}
	[\mathcal{C}_{DCBEF}] = \{x\in\Omega\,\, |\,\, \mathcal{C}_{DCBEF} \sim x\} = \{
	\overset{1}{\mathcal{C}_{DCBEF}},
	\overset{1}{\mathcal{C}_{CBEFD}},
	\overset{1/3}{\mathcal{C}_{FEDCB}}\}
	\end{equation}
	Here, the first two loops were found once, whereas the third cycle, $ \mathcal{C}_{FEDCB} $, was only found ``1/3-times'', meaning that it was one of three valid contractions of a given path (such as in the example of Fig.~\ref{fig:loop_algorithm}d). The cycles $ \mathcal{C}_{EFDCB} $ and $ \mathcal{C}_{BEFDC} $, which are also equivalent to $ \mathcal{C}_{DCBEF} $, are not found at all by the walker. In this more general case, the multiplicity of the equivalence class and therefore the weight of the associated polygon $ \mathcal{P}_{DCBEF} $ is given by the sum of the overnumbers, in this case $ w_{DCBEF} \equiv |[\mathcal{C}_{ABCD}]| = 1+1+1/3 = 7/3 < |\mathcal{C}_{DCBEF}|=5$.
	The polygons $ \mathcal{P}_i $ and corresponding weights $ w_i $ can be seen as the generalization of faces and their size in the planar case.

	Due to symmetry reasons, we always consider both right- and left-turning walkers. For graphs with no bond crossings this results in a double-counting which is accounted for by dividing all face weights by two. Note that for this reason we do not include inversions in the equivalence class definition (see Eq.~\ref{eq:equivalence_class_planar}). However, as soon as bonds cross, as for non-planar graphs, the left/right symmetry can be broken, as illustrated in Fig.\ref{fig:loop_algorithm}c. There, a right-turning walker starting from A to D finds the cycle ADBA, whereas the reverse search (a left-turning walker starting from D to A) finds the cycle DACD. Were the bond A-D instead outside the quadrilateral (denoted by the dashed circle segment), the symmetry would be retained, with both walkers traversing the cycle ADBA, and, respectively DABD, which are the same path. 
	
	Finally, note that the examples of Fig.~\ref{fig:loop_algorithm} are only meant to clarify the concepts introduced. For such small graphs the infinite face (\ie, external to all bonds) needs to be taken into account as well -- and the algorithm fails to do so properly when a site is present more than once in the path along the graph boundary (such as site C in Fig.~\ref{fig:loop_algorithm}b).
	For the large lattices under periodic boundary conditions we consider, though, a ``bulk" algorithm is fully adequate.
	
	We perform the search procedure using both right and left walkers, starting from every site and direction. The resulting set of polygons and corresponding weights is what defines our generalised dual tessellation
	\begin{align}
	T\equiv\{(\mathcal{P}_i,w_i), (\mathcal{P}_j,w_j), \ldots\}.
	\end{align}
	Every polygon covers a certain area of the spatial domain, as exemplified in Fig.~\ref{fig:loop_algorithm}d. The weight of the polygon is attributed to this area, as shown in Fig.~\ref{fig:dual_tessellation_schematic}. For planar graphs (Fig.~\ref{fig:dual_tessellation_schematic}b), the weights exactly represent the coordination numbers of the dual lattice sites, \ie $w$ are the number of vertices of the corresponding face. For dual tessellations of non-planar graphs (Fig.~\ref{fig:dual_tessellation_schematic}a) different cycles often overlap. The weights attributed to those regions are the sums of the weights of the individual overlapping polygons.
	More precisely, the total weight $W$ of a point $x$ in space is defined by
	\begin{equation}
	W(x) \equiv \sum_{i}w_i\quad |\quad x \text{ inside } \mathcal{P}([{\mathcal{C}_i}]),
	\end{equation}
	\ie the sum of the weights of polygons inside of which $x$ can be found. Examples of the dual tessellations (\ie weight fields $W(x)$) are shown in Fig.~\ref{fig:dual_tessellation_cg}.

\input{tikz_fig2}

%% file: tikz_fig1.tex
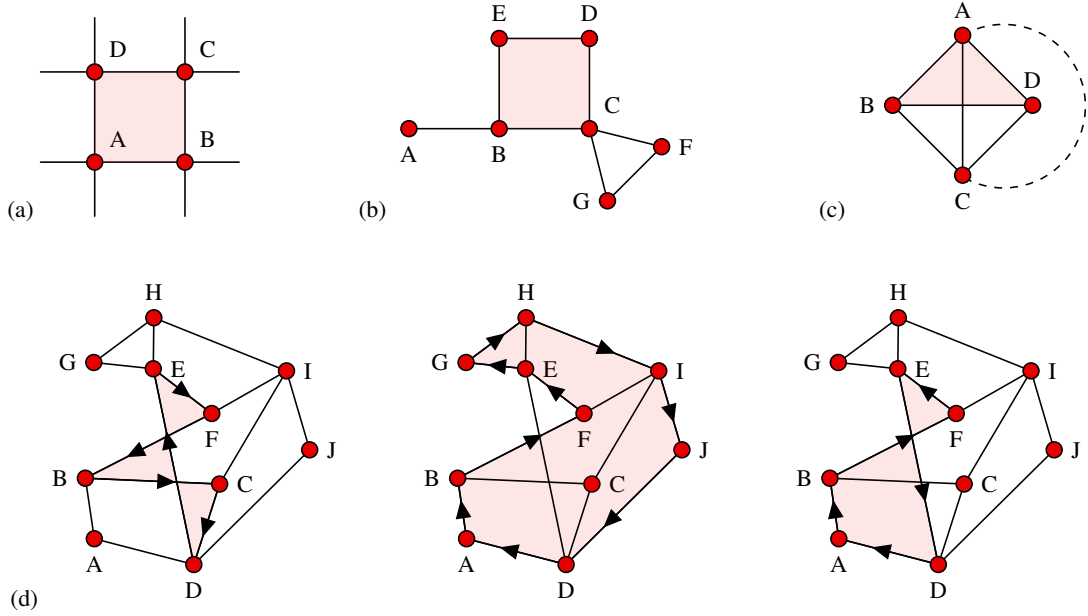
\begin{figure}[t]
	\centering
	\tikzset{mystyle/.style={draw, circle, semithick, fill=black!10!red, draw=black, text=white, font=\itshape, inner sep=0pt, minimum size=6pt}}
	\def\shift{5.0}
	\begin{small}
	(a)
	\begin{tikzpicture}[scale=1.2, line cap=round, line join=round, >=triangle 45,	every path/.style={black, semithick}]	
		\node[mystyle,label=above right:A] (A) at (0.0,0.0) {};
		\node[mystyle,label=above right:B] (B) at (1.0,0.0) {};
		\node[mystyle,label=above right:C] (C) at (1.0,1.0) {};
		\node[mystyle,label=above right:D] (D) at (0.0,1.0) {};

		\draw (A)--(B)--(C)--(D)--(A);
		\draw (0.0,-0.6) -- (A) -- (-0.6,0.0);
		\draw (1.0,-0.6) -- (B) -- (1.6,0.0);
		\draw (1.0,1.6)  -- (C) -- (1.6,1.0);
		\draw (-0.6,1.0) -- (D) -- (0.0,1.6);
	
		\fill[fill=black!10!red,fill opacity=0.1] (A.center) --(B.center) --(C.center) --(D.center) --cycle;;

	\end{tikzpicture}
	\hspace*{15mm}(b)
	\begin{tikzpicture}[scale=1.2, line cap=round, line join=round, >=triangle 45,	every path/.style={black, semithick}]	
		\node[mystyle,label=below:A] (A) at (-1.0,0.0) {};
		\node[mystyle,label=below:B]  (B) at (0.0,0.0) {};
		\node[mystyle,label=above right:C] (C) at (1.0,0.0) {};
		\node[mystyle,label=above:D] (D) at (1.0,1.0) {};
		\node[mystyle,label=above:E] (E) at (0.0,1.0) {};
		\node[mystyle,label=right:F] (F) at (1.8,-0.2) {};
		\node[mystyle,label=left:G]  (G) at (1.2,-0.8) {};
		
		\draw (A)--(B)--(E)--(D)--(C)--(F)--(G)--(C)--(B);
		
		\fill[fill=black!10!red,fill opacity=0.1] (B.center) --(C.center) --(D.center) --(E.center) --cycle;;
	
	\end{tikzpicture}
	\hspace*{15mm}(c)
	\begin{tikzpicture}[scale=0.93, line cap=round, line join=round, >=triangle 45,	every path/.style={black, semithick}]	
		\node[mystyle,label=above:A] (A) at (0,1) {};
		\node[mystyle,label=left:B] (B) at (-1,0) {};
		\node[mystyle,label=below:C] (C) at (0,-1) {};
		\node[mystyle,label=above:D] (D) at (1,0) {};
		
		\draw (A)--(B)--(C)--(D)--(A);
		\draw (A)--(C);
		\draw (B)--(D);
		
		\fill[fill=black!10!red,fill opacity=0.1] (A.center) --(D.center) --(B.center)--cycle;;
		
		\begin{scope}[on background layer]
			\draw[dashed] (A) arc (120:-120:11.7mm);
		\end{scope}
	
	\end{tikzpicture} \\\vspace*{7mm}(d)
	\begin{tikzpicture}[scale=0.7, line cap=round, line join=round, >=triangle 45,	every path/.style={black, semithick}]	
		\node[mystyle,label=below:A] (A) at (4.09,3.48) {};
		\node[mystyle,label=left:B]  (B) at (3.92,4.63) {};
		\node[mystyle,label=right:C] (C) at (6.47,4.52) {};
		\node[mystyle,label=below:D] (D) at (5.98,2.99) {};
		\node[mystyle,label=right:E] (E) at (5.21,6.71) {};
		\node[mystyle,label=below:F] (F) at (6.32,5.86) {};
		\node[mystyle,label=left:G]  (G) at (4.08,6.83) {};
		\node[mystyle,label=above:H] (H) at (5.22,7.68) {};
		\node[mystyle,label=right:I] (I) at (7.74,6.67) {};
		\node[mystyle,label=right:J] (J) at (8.18,5.17) {};

		\fill[fill=black!10!red,fill opacity=0.1] (E.center) --(F.center) --(B.center) --(C.center) --(D.center)--cycle;;
		
		\draw (B)--(C)--(D)--(E)--(G)--(H)--(I)--(F)--(B)--(A)--(D)--(J)--(I)--(C);
		\draw (H)--(E)--(F);
		
		\begin{scope}[semithick,decoration={markings,mark=at position 0.7 with {{\small \arrow{>}}}}]
			\path[draw,postaction={decorate}] (E)--(F);
			\path[draw,postaction={decorate}] (F)--(B);
			\path[draw,postaction={decorate}] (B)--(C);
			\path[draw,postaction={decorate}] (C)--(D);
			\path[draw,postaction={decorate}] (D)--(E);
		\end{scope}
	\end{tikzpicture}
	\hspace*{8mm}
	\begin{tikzpicture}[scale=0.7, line cap=round, line join=round, >=triangle 45,	every path/.style={black, semithick}]	
		\node[mystyle,label=below:A] (A) at (4.09,3.48) {};
		\node[mystyle,label=left:B]  (B) at (3.92,4.63) {};
		\node[mystyle,label=right:C] (C) at (6.47,4.52) {};
		\node[mystyle,label=below:D] (D) at (5.98,2.99) {};
		\node[mystyle,label=right:E] (E) at (5.21,6.71) {};
		\node[mystyle,label=below:F] (F) at (6.32,5.86) {};
		\node[mystyle,label=left:G]  (G) at (4.08,6.83) {};
		\node[mystyle,label=above:H] (H) at (5.22,7.68) {};
		\node[mystyle,label=right:I] (I) at (7.74,6.67) {};
		\node[mystyle,label=right:J] (J) at (8.18,5.17) {};
		
		\path[name path=cycle1] (F)--(E)--(G)--(H)--(I)--(J)--(D)--(A)--(B)--(F);
		\fill[fill=black!10!red,fill opacity=0.1] (F.center) --(E.center) --(G.center) --(H.center) --(I.center) --(J.center) --(D.center) --(A.center) --(B.center) --cycle;;
		
		\draw (B)--(C)--(D)--(E)--(G)--(H)--(I)--(F)--(B)--(A)--(D)--(J)--(I)--(C);
		\draw (H)--(E)--(F);
		
		\begin{scope}[semithick,decoration={markings,mark=at position 0.7 with {\arrow{>}}}]
			\path[draw,postaction={decorate}] (F)--(E);
			\path[draw,postaction={decorate}] (E)--(G);
			\path[draw,postaction={decorate}] (G)--(H);
			\path[draw,postaction={decorate}] (H)--(I);
			\path[draw,postaction={decorate}] (I)--(J);
			\path[draw,postaction={decorate}] (J)--(D);
			\path[draw,postaction={decorate}] (D)--(A);
			\path[draw,postaction={decorate}] (A)--(B);
			\path[draw,postaction={decorate}] (B)--(F);
		\end{scope}
	\end{tikzpicture}
	\hspace*{8mm}
	\begin{tikzpicture}[scale=0.7, line cap=round, line join=round, >=triangle 45,	every path/.style={black, semithick}]	
		\node[mystyle,label=below:A] (A) at (4.09,3.48) {};
		\node[mystyle,label=left:B]  (B) at (3.92,4.63) {};
		\node[mystyle,label=right:C] (C) at (6.47,4.52) {};
		\node[mystyle,label=below:D] (D) at (5.98,2.99) {};
		\node[mystyle,label=right:E] (E) at (5.21,6.71) {};
		\node[mystyle,label=below:F] (F) at (6.32,5.86) {};
		\node[mystyle,label=left:G]  (G) at (4.08,6.83) {};
		\node[mystyle,label=above:H] (H) at (5.22,7.68) {};
		\node[mystyle,label=right:I] (I) at (7.74,6.67) {};
		\node[mystyle,label=right:J] (J) at (8.18,5.17) {};

		\fill[fill=black!10!red,fill opacity=0.1] (F.center) --(E.center) --(D.center) --(A.center) --(B.center) --cycle;;
		
		\draw (B)--(C)--(D)--(E)--(G)--(H)--(I)--(F)--(B)--(A)--(D)--(J)--(I)--(C);
		\draw (H)--(E)--(F);
		
		\begin{scope}[semithick,decoration={markings,mark=at position 0.7 with {\arrow{>}}}]
			\path[draw,postaction={decorate}] (F)--(E);
			\path[draw,postaction={decorate}] (E)--(D);
			\path[draw,postaction={decorate}] (D)--(A);
			\path[draw,postaction={decorate}] (A)--(B);
			\path[draw,postaction={decorate}] (B)--(F);
		\end{scope}
	\end{tikzpicture}
	\end{small}
	\caption{(a) Section of a regular square lattice where the loop algorithm finds the usual face. (b) Illustration of the pruning process. (c) Simple example, where a bond, which breaks planarity, also breaks the symmetry between left- and right-turning walkers. (d) Non-unique contractions result in different loops.}
\label{fig:loop_algorithm}
\end{figure}

%% file: tikz_fig2.tex
\begin{figure*}
	\centering
		\tikzset{mystyle/.style={draw, circle, semithick, fill=black!10!red, draw=black, text=white, font=\itshape, inner sep=0pt, minimum size=6pt}}

	\begin{tikzpicture}[scale=3.5, semithick, line cap=round, line join=round, >=triangle 45,	every path/.style={black}]	
		\node[mystyle] (a1) at (0.6,1.2) {};
		\node[mystyle] (a2) at (1,1.8) {};
		\node[mystyle] (a3) at (1.5,0.9) {};
		\node[mystyle] (a4) at (1.2,1.6) {};
		\node[mystyle] (a5) at (1.2,0.8) {};
		\node[mystyle] (a6) at (2,0.8) {};
		\node[mystyle] (a7) at (2,1.6) {};
		\node[mystyle] (a8) at (0.8,1) {};
		\node[mystyle] (a9) at (1,1.2) {};
		\node[mystyle] (a10) at (1.6,1.2) {};
		\node[mystyle] (a11) at (1.8,0.6) {};
		\node[mystyle] (a12) at (1.2,0.6) {};
		\node[mystyle] (a13) at (0.6,1) {};
		\node[mystyle] (a14) at (0.4,0.6) {};
		\node[mystyle] (a15) at (0.8,0.6) {};
		\node[mystyle] (a16) at (3.59,0.68) {};
		\node[mystyle] (a17) at (3.59,0.98) {};
		\node[mystyle] (a18) at (3.34,1.13) {};
		\node[mystyle] (a19) at (3.08,0.98) {};
		\node[mystyle] (a20) at (3.08,0.68) {};
		\node[mystyle] (a21) at (3.34,0.53) {};
		\node[mystyle] (a22) at (3.74,1.24) {};
		\node[mystyle] (a23) at (3.49,1.39) {};
		\node[mystyle] (a24) at (2.84,1.08) {};
		\node[mystyle] (a25) at (2.79,1.48) {};
		\node[mystyle] (a26) at (3.09,1.58) {};
		\node[mystyle] (a27) at (2.99,1.38) {};
		\node[mystyle] (a28) at (3.39,1.68) {};
		\node[mystyle] (a29) at (2.69,0.78) {};
		\node[mystyle] (a30) at (2.89,0.58) {};
		
		\draw (a1)--(a2)--(a3)--(a1);
		\draw (a4)--(a5)--(a6)--(a7)--(a4);
		\draw (a8)--(a9)--(a10)--(a11)--(a12)--(a13)--(a14)--(a15)--(a8);
		
		\fill[color=black, fill=black!10!red,fill opacity=0.1] (a1.center)--(a2.center)--(a3.center)--(a1.center);
		\fill[color=black, fill=black!10!red,fill opacity=0.1] (a4.center)--(a5.center)--(a6.center)--(a7.center)--(a4.center);
		\fill[color=black, fill=black!10!red,fill opacity=0.1] (a8.center)--(a9.center)--(a10.center)--(a11.center)--(a12.center)--(a13.center)--(a14.center)--(a15.center)--(a8.center);
		
		\draw (a16)--(a17)--(a18)--(a19)--(a20)--(a21)--(a16);
		\draw (a17)--(a22)--(a23)--(a18);
		\draw (a24)--(a25)--(a26)--(a27)--(a18)--(a19)--(a24);
		\draw (a28)--(a23)--(a18)--(a27)--(a26)--(a28);
		\draw (a30)--(a29)--(a24)--(a19)--(a20)--(a30);
		\draw (a30)--(a21)--(a20)--(a30);
		
		\fill[color=black, fill=black!10!red,fill opacity=0.1] (a16.center)--(a17.center)--(a18.center)--(a19.center)--(a20.center)--(a21.center)--(a16.center);
		\fill[color=black, fill=black!10!red,fill opacity=0.1] (a17.center)--(a22.center)--(a23.center)--(a18.center);
		\fill[color=black, fill=black!10!red,fill opacity=0.1] (a24.center)--(a25.center)--(a26.center)--(a27.center)--(a18.center)--(a19.center)--(a24.center);
		\fill[color=black, fill=black!10!red,fill opacity=0.1] (a28.center)--(a23.center)--(a18.center)--(a27.center)--(a26.center)--(a28.center);
		\fill[color=black, fill=black!10!red,fill opacity=0.1] (a30.center)--(a29.center)--(a24.center)--(a19.center)--(a20.center)--(a30.center);
		\fill[color=black, fill=black!10!red,fill opacity=0.1] (a30.center)--(a21.center)--(a20.center)--(a30.center);
		
		\begin{small}
			\draw (1.73,1.44) node[anchor=center] {8/3};
			\draw (0.93,1.48) node[anchor=center] {7/3};
			\draw (0.6,0.82) node[anchor=center] {8};
			\draw (1.02,0.98) node[anchor=center] {8};
			\draw (1.08,1.12) node[anchor=center] {31/3};
			\draw (1.24,1.28) node[anchor=center] {5};
			\draw (1.27,1.12) node[anchor=center] {13};
			\draw (3.23,1.45) node[anchor=center] {5};
			\draw (3.53,1.20) node[anchor=center] {4};
			\draw (2.97,1.24) node[anchor=center] {6};
			\draw (3.31,0.86) node[anchor=center] {6};
			\draw (2.92,0.85) node[anchor=center] {5};
			\draw (3.09,0.605) node[anchor=center] {3};
			\draw (1.53,1.07) node[anchor=center] {32/3};
			\draw (0.5,1.76) node[anchor=center] {a)};
			\draw (2.63,1.76) node[anchor=center] {b)};
		\end{small}

	\end{tikzpicture}
	\caption{Schematic visualization of the dual tessellation. The numbers denote the corresponding weights of the polygons $ \mathcal{P} $ (generalized faces). In the example of a non-planar lattice (a) the polygons intersect and the weights are added in the overlapping regions.
	In the planar case (b) no overlaps exist and the weights reduce to the number of sites in the polygon. }
	\label{fig:dual_tessellation_schematic}		
\end{figure*}
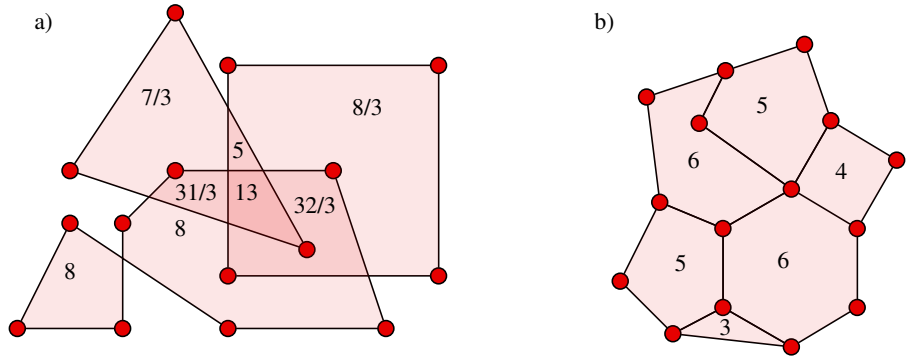